 \documentclass[final,5p,times,twocolumn]{elsarticle}
\usepackage{amsmath,amssymb,amsfonts}
\usepackage{slashed}
\usepackage{bbm}
\usepackage{nicefrac}
\usepackage{dsfont}
\usepackage{fixmath}
\usepackage{mathtools}

\usepackage{amssymb}

\newcommand\ii{\mathbbm{i}}
\DeclareMathOperator{\idm}{\mathds{1}}
\newcommand*{\field}[1]{\mathbb{#1}}%
\journal{Physics Letters B}

\begin{document}

\begin{frontmatter}



\title{A non-integrable quench from AdS/dCFT }

\author{Marius de Leeuw$^a$, Charlotte Kristjansen$^b$, and Kasper E. Vardinghus$^b$}

\address[label1]{School of Mathematics \& Hamilton Mathematics Institute\\
Trinity College Dublin \\
Dublin, Ireland}

\address[label2]{Niels Bohr Institute, Copenhagen University,\\ Blegdamsvej 17, 2100 Copenhagen \O, Denmark}
 


\begin{abstract}
We study the matrix product state which appears as the boundary state of the AdS/dCFT set-up where a probe D7 brane wraps two two-spheres stabilized by fluxes. 
The matrix product state plays a dual role, on one hand acting as a tool for computing one-point functions in a domain wall version of ${\cal N}=4$ SYM and on the other hand acting as the initial state in the study of quantum quenches of the Heisenberg spin chain.   We derive a number of selection
 rules for the overlaps between the matrix product state and the eigenstates of the Heisenberg spin chain  and  in particular demonstrate that the matrix product state does not fulfill a recently proposed integrability criterion. Accordingly, we find that the overlaps can not be expressed in the usual factorized determinant form. Nevertheless, we derive some exact results for one-point functions of simple operators and present a closed  formula for one-point functions of more general operators  in the limit of large spin-chain length. 
\end{abstract}


\end{frontmatter}



\section{Introduction}

Exact results for overlaps between states in integrable spin chains have important applications in the calculation of correlation functions in supersymmetric gauge theories as well as in 
the study of quantum quenches in statistical physics.  
Recently, especially overlaps between Bethe eigenstates and  matrix product states
have attracted attention. From the point of view of the AdS/dCFT correspondence,  overlaps between Bethe eigenstates and
specific matrix product states  encode information about one-point functions
 in domain wall versions of ${\cal N}=4$ SYM 
 theory~\cite{deLeeuw:2015hxa,Buhl-Mortensen:2015gfd,deLeeuw:2016umh,deLeeuw:2016ofj,deLeeuw:2018mkd} 
 and in statistical physics the same matrix product states play the role of the initial state of a
quantum quench~\cite{Mestyan:2017xyk,Piroli:2018ksf,Piroli:2018don}.

Interestingly, all spin chain states $|\Psi\rangle$ for which it has been possible to write the overlap with the Bethe eigenstates in a closed form 
have been characterized by being annihilated by the entire tower of parity odd conserved charges of the chain.
Furthermore,
for all of these cases the annihilation of the state by the odd charges could be used to show that the overlaps with Bethe eigenstates were only non-vanishing for Bethe states with  paired roots\footnote{States with paired roots are states for
which the roots take the form $\{u_i,-u_i\}\bigcup S_{u}$, where $q_{2n+1}(u)=0 $ for 
$u\in S_{u}$. For the SU(2) Heisenberg spin chain that we consider in the present letter, $S_{u}=\emptyset$,  but
for spin chains with nested Bethe ans\"{a}tze such as the SU(3) or the SO(6) spin chain there can be a single root a 
zero~\cite{deLeeuw:2016umh,deLeeuw:2018mkd}.} and finally  the overlaps took a factorized form with
the Gaudin norm matrix, $G$ \cite{Gaudin,Korepin:1982ej},
playing a prominent role. More precisely, for Bethe states with paired roots the determinant of the Gaudin matrix factorizes as~\footnote{For a detailed explanation of how this happens for a model with a nesting we refer 
to~\cite{deLeeuw:2016umh}.}
\begin{equation}
\det G=\det G_+ \det G_-,
\end{equation}
and the normalized overlap takes the (schematic) form
\begin{equation}
\frac{\langle \Psi|\, {\bf{u}}\rangle}{\,\,\,\,\langle {\,\bf{u}\,}|\,{\bf{u}}\,\rangle^{1/2}}=\prod_i f({u_i}) \sqrt{\frac{\det{G_+}}{\det{G_-}}}.
\label{closed}
\end{equation}
These observations lead the authors of~\cite{Piroli:2017sei} to suggest that 
matrix product
states should be denoted as integrable when annihilated by all odd charges of the spin chain and in that case would play a role
analogous to that of the integrable boundary states of Zamolodnikov for continuum quantum field theories~\cite{Ghoshal:1993tm}.  Furthermore, in~\cite{Pozsgay:2018dzs} integrable matrix product states were related to novel types of  solutions 
to the twisted Boundary Yang-Baxter equations, carrying extra
internal degrees of freedom, .

  Note, however, that the notion of integrability of a matrix product state has (so far) not been used neither to prove the existence of, nor to derive a closed expression for the overlaps with the Bethe eigenstates. Furthermore, it is not excluded that a matrix
product state which is not integrable in the sense  above could have a closed formula describing its overlaps with the
Bethe eigenstates and finally the integrability criterion only directly applies to spin chains for which the conserved charges
can be defined to have a specific parity.

The approach of using matrix product states in the calculation of one-point functions in defect versions of ${\cal N}=4$ SYM theory was introduced in~\cite{deLeeuw:2015hxa,Buhl-Mortensen:2015gfd}  where the field theory was taken to have gauge groups
of different rank, $U(N)$ and $U(N-k)$, on the two sides of a co-dimension one defect~\cite{Karch:2000gx,Gaiotto:2008sa}. 
This domain wall set-up has a dual string theoretical description as  a D3-D5 probe brane system where the D5 probe has geometry
$AdS_4 \times S^2$ and where  there are $k$ units
of magnetic flux through the $S^2$~\cite{Karch:2000gx,Constable:1999ac,Constable:2001ag}.  
In this case it was possible to find a closed expression of the form~(\ref{closed}) for the one-point functions of all scalar operators which involved finding a closed expression for the overlap between a matrix product state and a Bethe eigenstate of the SO(6) spin chain~\cite{deLeeuw:2018mkd}.

The approach was pursued for a different D3-D5 based
defect version of ${\cal N}=4$ SYM in~\cite{Widen:2018nnu}, 
namely that constructed from the $\beta$-deformed theory . 
These investigations did not reveal a closed formula for the one-point functions. In this case, neither the Hamiltonian, nor the
higher conserved charges of the associated integrable spin have a definite parity and thus the integrability criterion above does not immediately apply. 
  
There exists another AdS/dCFT set-up which is very similar to the  D3-D5 probe brane system and which
also leads to a domain wall version of ${\cal N}=4$ SYM theory, namely a D3-D7 probe brane system, likewise with background
gauge field flux. The D3-D7 probe brane set-up comes in two different versions corresponding to two different probe brane
embeddings with respectively SO(5) and SO(3)$\times$SO(3) symmetry~\cite{Myers:2008me,Bergman:2010gm,Kristjansen:2012tn}.
In  the SO(5) symmetric
case  the matrix product state of relevance for the computation of scalar one-point functions belongs to the integrable class in the sense above~\cite{deLeeuw:2018mkd}.  We note, however, that at the present moment
a closed expression for the one-point functions is not known~\cite{deLeeuw:2016ofj}.

In this paper we will study the matrix product state that encodes the one-point functions of the SO(3)$\times$SO(3) symmetric
D3-D7 probe brane system and show that as opposed to its above mentioned relatives it does not qualify as an integrable boundary state. In accordance with this we find that the one-point functions can not be written in the form of \eqref{closed} and
no indication of an alternative closed formula in terms of determinants was observed. Nevertheless, we are still able to extract non-trivial exact information about the one-point functions of the corresponding dCFT.

Let us mention that very recently matrix product states have made their appearance in the calculation of three-point functions in ${\cal N}=4$ SYM theory involving two determinant operators and one single trace non-protected operator~\cite{Jiang:2019xdz}. This is very natural  as the dual string theory computation is very similar to the one required for the
computation of one-point functions~\cite{Bissi:2011dc} with the parameter describing the background gauge-field flux being replaced by the angular momentum of a giant graviton.

Our letter is organized as follows. In section~\ref{MPS_intro} we introduce the relevant matrix product state and sketch its role in the calculation of one-point functions.  
We shall be brief regarding this point and refer to~\cite{deLeeuw:2017cop,Marius_review} for details. Subsequently, in section~\ref{Integrability_test}, we investigate the action of the simplest odd charge on the matrix product state and derive a number of selection rules for the one-point functions of the corresponding dCFT.  In section~\ref{exact} we present  a few
exact results for  one-point functions of simple operators and, in particular, we quantify the deviation of the results from the
formula~(\ref{closed}). Finally,  in section~\ref{Large-L} we present a closed formula for the one-point functions in the limit of large-$L$, where $L$ is the number of fields in the operator considered, respectively the length of the spin chain involved.
Section~\ref{Conclusion} contains our conclusion.

\section{The Matrix Product State \label{MPS_intro}}
The classical equations of motion of ${\cal N}=4$ SYM admit a fuzzy funnel solution where for $x_3>0$
the six scalar fields take the values~\cite{Kristjansen:2012tn}
\begin{align}
 \label{eq:classical-solution-so3-so3}
 \begin{split}
 \phi^{cl}_i(x) &= - \frac{1}{x_3} \left( t_i^{k_1} \otimes {\idm}_{k_2} \right) \oplus 0_{N-k_1 k_2} \quad \text{for} \quad i = 1, 2, 3, \\
 \phi^{cl}_i(x) &= - \frac{1}{x_3} \left({\idm}_{k_1} \otimes t_{i-3}^{k_2} \right) \oplus 0_{N-k_1 k_2} \quad \text{for} \quad i = 4, 5, 6,
 \end{split}
\end{align}
while the fermionic fields as well as the gauge fields vanish, and where for $x_3<0$
 all fields carry a SU($N-k_1k_2$) representation and  vanish in the classical limi.   Here, the matrices $t_i^{k_a}$ constitute a $k_a$-dimensional irreducible representation of SU(2).
 This solution realizes a domain wall which separates a region $(x_3<0)$ where the 
field theory has gauge group SU($N-k_1 k_2$) from a region $(x_3>0)$ where the theory has gauge group  SU($N$), broken by the vevs.
We shall be interested in studying the tree-level one-point functions in the SU(2) sub-sector of conformal operators, built from the complex fields
$Z$ and $X$ defined by
\begin{equation}
X=\phi_1+\ii \phi_4, \hspace{0.5cm} Z=\phi_2+\ii\phi_5,
\end{equation}
and described by a certain eigenstate $|\{u_i\}\rangle\equiv|{\bf{u}}\rangle$ of the integrable Heisenberg spin chain where the $u_i$ are the
corresponding Bethe roots~\cite{Minahan:2002ve}. Already in~\cite{Kristjansen:2012tn} a closed expression for the overlap of the vacuum state with the Bethe eigenstates was found and matched to a string theory result. This match between gauge and string theory was recently extended to the next to leading
order in~\cite{Grau:2018keb}.
 Here we will deal with excited states corresponding to 
 non-protected operators in the field theory.  Computing the tree-level value of the one point functions, which amounts to inserting the classical values for the fields in the expressions for the conformal operators,   can be implemented by means of  the
 following  matrix product
state
\begin{equation}
\label{MPS}
 \langle \text{MPS}_{(k_1,k_2)}(\alpha)|  =
 \text{tr}\prod_{n=1}^L\Big(\,\langle \, \uparrow|_n\, \otimes T_1^{(k_1,k_2)}(\alpha)+\langle\,\downarrow |_n \,\otimes T_2^{(k_1,k_2)}(\alpha)\Big),
\end{equation}
where
\begin{equation}
T_i^{(k_1,k_2)}(\alpha)=  t_i^{k_1} \otimes {\idm}_{k_2} +\alpha  {\idm}_{k_1} \otimes  t_i^{k_2}.
\end{equation}
The introduction of  the parameter $\alpha$ allows us to write the commutation relation for the $T$ matrices as
\begin{equation}
 \left[T_i^{(k_1,k_2)}(\alpha),T_j^{(k_1,k_2)}(\beta)\right]=\ii\varepsilon _{ijk}T_k^{(k_1,k_2)}(\alpha \beta).
\end{equation}
The parameter $\alpha$ also allows us to interpolate between various models. The case $\alpha=\pm\ii$ will be relevant for the computation of the D3-D7 one-point functions, while the cases $\alpha=0,\pm1$ are related to the D3-D5 probe brane matrix product state. 

More precisely, the one-point functions of interest for the D3-D7 brane case can be expressed as 
\begin{equation}
\langle {\cal O}_{L}\rangle= \left(\frac{8\pi ^2}{\lambda }\right)^{\frac{L}{2}}L^{-\frac{1}{2}}\frac{C_{k_1,k_2}}{x_3^{\,L}},
\end{equation}
where
\begin{equation}\label{genericCso6}
 C_{k_1,k_2}=
 \frac{\left\langle \,{\bf{u}}\,\right.\!\!\left|\vphantom{\bf u}{\rm MPS}_{(k_1,k_2)}(\alpha=\ii) \right\rangle}{\left\langle \,{\bf{u}}\, \right.\!\!\,\left|\, {\bf{u}}\,  \right\rangle^{\frac{1}{2}}}.
\end{equation}
 The case $\alpha=0$ is trivially related to the D3-D5 probe brane matrix product state but all other cases differ from the latter
in a crucial manner. In particular,  one can easily convince oneself, that as opposed to what  was the case
in the D3-D5 set-up~\cite{Buhl-Mortensen:2015gfd} there is no recursive relation which connects matrix product states with different bond dimensions to each other. 
We notice, however,  that by setting $k_1=1$ or $k_2=1$ we recover the matrix product state of relevance for the D3-D5 probe brane
set-up~\cite{deLeeuw:2015hxa,Buhl-Mortensen:2015gfd}. 
\section{Integrability Test and  Selection Rules\label{Integrability_test}}
 
Using the explicit expression for the simplest odd charge, $Q_3$, of the SU(2) Heisenberg spin chain (with periodic boundary
conditions)
\begin{equation}
Q_3= \sum_{n=1}^L [P_{n,n+1},P_{n+1,n+2}],
\end{equation}
where $P$ is the permutation operator,
one finds that
\begin{equation}
Q_3 | \text{MPS}_{(k_1,k_2)}(\alpha)\rangle \neq 0,
\end{equation}
for $L\geq 12$ and for all values of $\alpha, k_1$ and $k_2$ except the trivial ones where the matrix product
state is related to the matrix product state of the D3-D5 probe brane set-up.  Hence the 
state~(\ref{MPS}) does not belong to the class of  matrix product states denoted as integrable and in nothing prevents Bethe eigenstates with unpaired roots from having a 
non-vanishing overlap with this state. Indeed, one easily finds by explicit computation examples of  Bethe eigenstates with un-paired roots and with non-vanishing overlap with the matrix
product state~(\ref{MPS}). Such Bethe eigenstates are first encountered for $L=12, M=6$, where $M$ is the number of excitations. Furthermore,  even for Bethe states with only paired roots explicit computations of overlaps 
have not revealed a closed formula for the overlaps.
However, one can still derive a number of exact results. In order to do so  it is useful to start by deriving a set of selection rules.  First, we notice that 
in order for a Bethe state to have a non-vanishing overlap with the matrix product state~(\ref{MPS}) it needs to have an
even length and an even number of excitations.  This result follows from the SU(2) algebra having the following authormorphisms~\cite{deLeeuw:2015hxa}
\begin{eqnarray}
Ut_1 U^{-1}=t_1,&\,\,& Ut_{2,3} U^{-1}=-t_{2,3}, \\
 Vt_3 V^{-1}=t_3,&\,\, &Vt_{1,2} V^{-1}=-t_{1,2}, 
\end{eqnarray}
with $U$ and $V$ unitary matrices, which naturally lift to the algebra of the $T_i^{(k_1,k_2)}$ so that for inst.
\begin{equation}
(U_1 \otimes U_2)\, T_1^{(k_1,k_2)}(\alpha)\, (U_1^{-1} \otimes U_2^{-1})= T_1^{(k_1,k_2)} (\alpha).
\end{equation}
Secondly, we have the relation
\begin{equation}
(\idm\otimes V)\, T_{1,2}^{(k_1,k_2)}(\alpha)\,  (\idm\otimes V^{-1})= T_{1,2}^{(k_1,k_2)}(-\alpha),
\end{equation}
and in addition there exists an invertible matrix $S$, a so-called shuffle matrix, which interchanges the factors in the direct matrix product, i.e.
\begin{equation}
S\,(\idm\otimes \,t_i)\, S^{-1} = t_i \otimes \idm,
\end{equation}
which for $k_1=k_2=k$ implies 
\begin{equation}
S\,T_{1,2}^{(k,k)}(\alpha)\, S^{-1} = \alpha T_{1,2} ^{(k,k)}(1/\alpha).
\end{equation}
Together these relations give for $k_1=k_2=k$  
\begin{eqnarray}
\text{Tr} \,(T_{s_1}(\alpha)\ldots T_{s_L}(\alpha))&=& \alpha^L \, \text{Tr} \,(T_{s_1}(1/\alpha)\ \ldots T_{s_L}(1/\alpha))\nonumber\\
&=&\text{Tr}\, (T_{s_1}(-\alpha)\ldots T_{s_L}(-\alpha)), \nonumber
\end{eqnarray}
with $s_i\in\{1,2\}$. Thus for $\alpha=\pm \ii$ we need that $\ii^L=1$, i.e. $L/4\in \field{N}$ in order for the overlap not to vanish.

Finally, we are only
interested in cyclically symmetric Bethe eigenstates, i.e. Bethe eigenstates with total momentum zero, as only such states can
represent a single trace operator of ${\cal N}=4$ SYM. We note, however, that due to the cyclicity of the matrix product state, its overlap
with Bethe eigenstates is vanishing for non-cyclic states.

\section{Exact Results \label{exact}}

As explained in~\cite{deLeeuw:2015hxa} the coordinate space Bethe ansatz provides an explicit expression for the Bethe
eigenstates which is useful for the calculation of the overlaps. More precisely, we have for the overlap of the matrix product state with a Bethe eigenstate with $M$ excitations
\begin{equation} 
\langle \text{MPS}|\,{\bf{u}}\,\rangle = {\cal N} \sum_{\sigma\in S_M} A_{\sigma} \sum_{1\le n_1< \ldots < n_M\leq L} \prod_{j=1}^M x_{\sigma_j}^{n_j}
\,\langle \text{MPS}|\{n_i\}\rangle, 
\label{overlap}
\end{equation}
where $A_{\sigma}$ is a product of two particle scattering matrices corresponding to the permutation $\sigma$ and where
\begin{equation}
x_j=\frac{u_j+\frac{\ii}{2}}{u_j-\frac{\ii}{2}}.
\end{equation} 
Furthermore,
\begin{align} 
\langle\text{MPS}|\{n_i\}\rangle &= \text{tr}(T_1\ldots T_1T_2T_1\ldots T_1T_2 T_1\ldots T_1) \label{tracefactor}
\end{align}
where the $M$ generators of type $T_2$ are located at the sites $n_1,\ldots, n_M$. Finally ${\cal N}$ is a normalization constant
in the form of a phase which we will choose so that the one-point function coefficient $C_{k_1,k_2}$ is real and positive. For
details we refer to~\cite{deLeeuw:2015hxa}.  We note that due to the tensor structure in the matrix product state all trace
factors of the type~(\ref{tracefactor}), even that corresponding to the vacuum, involve  binomial sums~\cite{Grau:2018keb}.

 By means of the relation~(\ref{overlap}) we can evaluate the overlap between the matrix product state and the two-excitation
 state for any value of $\alpha$. For $k_1=k_2=2$ the resulting trace factor \eqref{tracefactor} can be simplified and evaluated explicitly\footnote{Here we are excluding the cases $\alpha=\pm 1$. For a discussion of these, see below.} 
 \begin{align} 
\langle\text{MPS}|\{n_i\}\rangle 
& = \bigg[
\Big(\frac{\alpha +1}{\alpha -1}\Big)^{\sum_i (-1)^i n_i} \Big(\frac{\alpha-1}{2}\Big)^{L} + 
\Big(\frac{\alpha -1}{\alpha +1}\Big)^{\sum_i (-1)^i n_i} \Big(\frac{\alpha +1}{2}\Big)^{L}
 \bigg]   \nonumber \\
& \times  \frac{2}{(\alpha^2-1)^{\frac{M}{2}}}\sum_{m=0}^\frac{M}{2}  \alpha^{2m} \sum_{\substack{A\subset\{1,2,\ldots,M\} \\ |A| = 2m}} (-1)^{\sum_i A_i} (-1)^{\sum_i n_{A_i}} .
\end{align}
From, this we can deduce the result for the overlap for  $k_1=k_2=M=2$, which reads
 \begin{align}
&\frac{\langle \textup{MPS}_{2,2}(\alpha) \vert \,{\bf{u}}\, \rangle}{\langle \,{\bf{u}}\,\vert \,{\bf{u}}\, \rangle^{1/2}} = \frac{u \sqrt{u^2 + \frac{1}{4}}}{2^{L-1}} 
\sqrt{\frac{L}{L-1}}\, \times \\
&\Bigg[\frac{\alpha  \left((\alpha -1)^{L-1}+(\alpha +1)^{L-1}\right)}{u^2+ \frac{1}{4 \alpha ^2}}-\frac{(\alpha -1)^{L-1}-(\alpha +1)^{L-1}}{u^2+ \frac{\alpha^2}{4}}\Bigg]. \nonumber
\end{align}
For $\alpha=\ii$ this reduces to
\begin{equation}\label{two-excitation-state}
\frac{\langle \textup{MPS}_{2,2}(\pm i) \vert \,{\bf{u}} \,\rangle}{\langle\, {\bf{u}}\, \vert \,{\bf{u}}\, \rangle^{1/2}}
=\frac{1}{2^{\frac{L}{2}-2}}\sqrt{\frac{L}{L-1}}  \frac{u \sqrt{u^2 + \frac{1}{4}}}{u^2 - \frac{1}{4}}.
\end{equation}
Furthermore, for $\alpha=0$ the overlap is proportional to the overlap with the D3-D5 matrix product state for $k=2$. Finally for
$\alpha=\pm 1$ the tensor product of the two two-dimensional SU(2) representations decompose as $3\oplus 1$ and
the overlap becomes proportional to the overlap with the D3-D5 matrix product state for $k=3$. 

For four excitations the result is non-zero even when the rapidities are unpaired. It can be computed exactly, but it is very lengthy. For the special where the rapidities are paired, the result can drastically simplifies
\begin{eqnarray}
\lefteqn{
\frac{\langle \textup{MPS}_{2,2}(\pm\ii)\vert \,{\bf{u}}\, \rangle}{\langle\, {\bf{u}}\, \vert \, {\bf{u}} \,\rangle^{1/2}} = \frac{1}{2^{\frac{L}{2}-2}} \Biggl[ \prod_{i=1}^2 \frac{u_i \sqrt{u_i^2 + \frac{1}{4} } }{ u_i^2 - \frac{1}{4} } \Biggr] \sqrt{ \frac{\det G^+}{\det G^-}} 
\nonumber }\\
&&+\frac{L}{2^{\frac{L}{2}+3} \sqrt{\det G}} \, \Biggl[ \prod_{i=1}^2 \frac{u_i }{ \bigr(u_i^2 - \frac{1}{4} \bigl)^2 \sqrt{u_i^2 + \frac{1}{4}}} \Biggr] \times \nonumber
\\
&&\frac{(u_1^2-u_2^2)^2 (-1-4u_1^2 -4u_2^2 + 48u_1^2u_2^2)}{(1+(u_1-u_2)^2) (1 + (u_1 +u_2)^2)\bigl(u_1^2u_2^2-\frac{1}{16}\bigr)}. \label{Four-excitation-state}
\end{eqnarray}
Here, in the first line we have singled out the "integrable" piece of the overlap, cf.~eqn.~(\ref{closed}). We notice that the remaining  part of the expression is subdominant in the
limit $L\rightarrow \infty$ behaving as ${\cal O}(\frac{1}{L})$ as opposed to the ${\cal O}(1)$ behaviour of the first term. Moreover, we see that even for states with paired rapidities, the overlap can not be written in the form of \eqref{closed}. We have not been able to find any determinant type formula which reproduces this result.

It is likewise possible to find the overlaps~(\ref{two-excitation-state}) and~(\ref{Four-excitation-state}) for higher values of $k_1$ and $k_2$ but unlike what was the case for the
D3-D5 set-up there does not seem to exist a recursion relation that relates the overlaps for different values of $k_1$ and $k_2$ which can be traced back  to the lack of a recursive relation between the matrix product states corresponding to different values of $(k_1,k_2)$. For $M=2$ and general values of $k_1$ and $k_2$ we find
\begin{align}
&\frac{\langle \textup{MPS}_{2,2}(\pm i) \vert \,{\bf{u}} \,\rangle}{\langle\, {\bf{u}}\, \vert \,{\bf{u}}\, \rangle^{1/2}}=u \sqrt{u^2+\frac{1}{4}}  \sqrt{\frac{L}{L-1}} 
\Bigg\{ \\
&\sum_{n=-\frac{k_1}{2}}^{\frac{k_1}{2}}
\sum_{m = -\frac{k_2-1}{2}}^{\frac{k_2-1}{2}}  
\bigg[n^2-\frac{k_1^2}{4}\bigg] 
\frac{(\alpha  m+n+\frac{1}{2})^{L-1}}{( \alpha m+n)^2+u^2}  +  \\
&\sum_{m=-\frac{k_2}{2}}^{\frac{k_2}{2}}\sum_{n = -\frac{k_1-1}{2}}^{\frac{k_1-1}{2}} 
\bigg[m^2-\frac{k_2^2}{4}\bigg]  
\alpha^L\frac{ (\alpha^{-1} n + m +\frac{1}{2})^{L-1}}{(\alpha^{-1} n+  m)^2+ u^2} \Bigg\}.
\end{align}

 As in~\cite{deLeeuw:2015hxa,Grau:2018keb} it is possible to extract the leading $k_1,k_2$ limit of the overlap, a quantity
 which is of relevance for comparison with the string theory side, cf.~\cite{Nagasaki:2012re,Kristjansen:2012tn}. 

For two excitations we find 
\begin{align}\label{eq:largeKM2}
&\frac{\langle \textup{MPS}_{2,2}(\alpha) \vert \,{\bf{u}}\, \rangle}{\langle \,{\bf{u}}\, \vert \,{\bf{u}}\,\rangle^{1/2}} =
\frac{2^{1-L}}{L(L-2)(L-3)} \sqrt{\frac{L}{L-1}} u \sqrt{u^2+\frac{1}{4}}\times \nonumber\\
&
\qquad\Bigg[(\alpha ^2+1)\frac{(k_1+ \alpha k_2)^L - (k_1-\alpha  k_2)^L}{\alpha } +\\
&\qquad
L
\frac{(k_1-\alpha ^3 k_2)(k_1-\alpha k_2)^{L-1} -(k_1 + \alpha ^3 k_2)(k_1+\alpha k_2)^{L-1}}{\alpha }
\Bigg].\nonumber
\end{align}
This greatly simplifies when $\alpha=\ii$ and we get
\begin{align}\label{eq:largeKM2i}
&\frac{\langle \textup{MPS}_{2,2}(\ii) \vert \,{\bf{u}}\, \rangle}{\langle \,{\bf{u}}\, \vert \,{\bf{u}}\,\rangle^{1/2}} =
\frac{2^{1-L}(k_1^2+k_2^2)}{(L-2)(L-3)} \sqrt{\frac{L}{L-1}} u \sqrt{u^2+\frac{1}{4}}\times \nonumber\\
&\qquad
\bigg[(k_1-i k_2)^{L-2} -(k_1+i k_2)^{L-2}
\bigg].\nonumber
\end{align}
Notice that the one-point function scales as $k^L$, whereas in the D3-D5 set-up the leading $k$ behaviour is of order $k^{L-1}$ for $M=2$. In fact, sending $k_1\rightarrow\infty$ while keeping $k_2$ finite should yield the usual D3-D5 result. Indeed, imposing this limit, we see that \eqref{eq:largeKM2} vanishes and we find that the subleading term reproduces the results from \cite{deLeeuw:2015hxa}.

\section{The Large-L Limit \label{Large-L}}
The large-$L$ limit of the overlaps is of relevance both for studying quantum quenches in the thermodynamical limit and for comparing with semi-classical string theory in the AdS/dCFT set-up~\cite{Buhl-Mortensen:2015gfd}.
In order to study the large-$L$ limit
of the overlaps, we notice that provided the trace factor $\langle \text{MPS}|\{n_i\}\rangle$ in eqn.~(\ref{overlap}) is exponential
in the $n_i$ the sum over the $n_i$ becomes geometrical 
and can be carried out explicitly. We observe that 
the leading $L$ contribution originates from terms in the sum over $n_i$ for which
paired rapidities are next to each other with each pair giving rise to a factor of $L$.  
 In particular, for $M$ excitations, states with all roots paired have an overlap with the matrix product state which behaves as $L^{M/2}$ for large-$L$,
whereas states with fewer paired roots have overlaps which scale with a lower power of $L$.

 We can use the observation above to facilitate the extraction of the large-$L$ behaviour of the overlaps.  Namely,  to determine the
 pre-factor of the leading $L$ term  we can
truncate the sum over permutations in~(\ref{overlap}) to only those which keep the  paired roots  next to each other. For $M$ excitations and
for states with only paired roots this reduces the number of
permutations in the sum from $M!$ to $2^{M/2}(M/2)! $. Based on explicit computations of the overlaps between the matrix product
state and Bethe states with  $M=2,4,6$ and only paired roots  we find the following expression for the large-$L$ contribution
to the one-point function for $k_1=k_2=2$
\begin{equation}
\begin{aligned}
\frac{\langle \textup{MPS}_{2,2}(\pm\ii) \vert \,{\bf{u}}\, \rangle }{\langle \,{\bf{u}}\, \vert \,{\bf{u}}\,\rangle^{1/2}}   =  \frac{1}{2^{\frac{L}{2}-2}} \prod_{i=1}^{\nicefrac{M}{2}} \frac{u_i \sqrt{u_i^2 + \frac{1}{4}}}{u_i^2 - \frac{1}{4}} +\mathcal{O}\bigl(\tfrac{1}{L} \bigr). \end{aligned}
\end{equation}
We notice that for $M=2$ there is no ${\cal O}\left(\frac{1}{L}\right)$ correction term, cf.~eqn. (\ref{two-excitation-state}).

\section{Summary and Conclusion \label{Conclusion}}

With the present investigations we have completed the analysis of the integrability structure of matrix product states of relevance for one-point functions in  defect versions of  (non-deformed) ${\cal N}=4$ SYM based on probe-brane set-ups with
fluxes. The supersymmetric D3-D5 probe brane set-up lead to a matrix product state fulfilling the integrability criterion 
of~\cite{Piroli:2017sei} and a closed formula for all scalar one-point functions of the dCFT could be derived~\cite{deLeeuw:2018mkd}. The two non-supersymmtric  D3-D7 probe brane set-ups have different behaviours. In the SO(5) symmetric case the relevant matrix product state is integrable in the sense that it is annihilated by all the odd charges of the
SO(6) spin chain but a closed expression for the one-point functions has so far not been found. For the D3-D7 set-up with 
SO(3)$ \times$SO(3) symmetry, studied here, we found that the matrix product state did not fulfill the proposed integrability criterion.  We mention, however, that for this case, as for the D3-D5 probe brane set-up, we still
have a complete match between one-point functions of chiral primaries computed in respectively string and gauge theory to two leading orders in a  double scaling limit~\cite{Grau:2018keb}.\footnote{The similar computation has not been carried out for
the SO(5) symmetric set-up.} Generalizing the string computation to non-protected operators constitutes an interesting
open problem in both models.

 Despite the lack of integrability indicators we were able to extract from our data a closed formula for  the large-$L$ limit of the one-point functions of operators corresponding to
 Bethe eigenstates with paired roots. The study of the large-$L$ limit of the overlaps was facilitated
by the observation that only a subset of the permutations appearing in the expression for coordinate space Bethe eigenfunctions
would contibute in this limit. This observation may  prove useful for the study of the large-$L$ limit of other similar overlap problems. 

Finally, let us mention that while the considerations in the present paper are mainly relevant for tree-level one-point functions,
the perturbative framework for calculating one-point functions at higher loop order for the here considered SO(3)$\times$SO(3) symmetric defect version of ${\cal N}=4$ SYM was set up in~\cite{Grau:2018keb}, generalizing the ideas 
of~\cite{Buhl-Mortensen:2016pxs,Buhl-Mortensen:2016jqo}.
In particular, using the framework of~\cite{Grau:2018keb} it is possible to compute the one-loop correction to the one-point functions given in 
eqns.~(\ref{two-excitation-state}) 
and~(\ref{Four-excitation-state}) and to test if tree-level and one-loop results are related via a simple flux factor as it was the case for the
D3-D5 probe brane set-up~\cite{Buhl-Mortensen:2017ind}.

 \vspace{0.2cm}

\section*{Acknowledgments}
We thank Asger Ipsen, Georgios Linardopoulos, Matthias Volk,  Matthias Wilhelm and  Konstantin Zarembo for useful discussions.
Furthermore, We thank Jan Ambj\o rn for giving us access to his computer system.
C.K.\ and K.E.V.\ were supported by DFF-FNU through grant number DFF-4002-00037. MdL was supported by SFI and the Royal Society for funding under grants UF160578 and RGF$\backslash$EA$\backslash$180167. 

\vspace*{0.5cm}




 \bibliographystyle{elsarticle-num}
 \bibliography{letter.bib}

\begin{thebibliography}{10}
\expandafter\ifx\csname url\endcsname\relax
  \def\url#1{\texttt{#1}}\fi
\expandafter\ifx\csname urlprefix\endcsname\relax\def\urlprefix{URL }\fi
\expandafter\ifx\csname href\endcsname\relax
  \def\href#1#2{#2} \def\path#1{#1}\fi

\bibitem{deLeeuw:2015hxa}
M.~de~Leeuw, C.~Kristjansen, K.~Zarembo, {One-point functions in defect CFT and
  integrability}, JHEP 08 (2015) 098.
\newblock \href {http://arxiv.org/abs/1506.06958} {\path{arXiv:1506.06958}},
  \href {http://dx.doi.org/10.1007/JHEP08(2015)098}
  {\path{doi:10.1007/JHEP08(2015)098}}.

\bibitem{Buhl-Mortensen:2015gfd}
I.~Buhl-Mortensen, M.~de~Leeuw, C.~Kristjansen, K.~Zarembo, {One-point
  functions in AdS/dCFT from matrix product states}, JHEP 02 (2016) 052.
\newblock \href {http://arxiv.org/abs/1512.02532} {\path{arXiv:1512.02532}},
  \href {http://dx.doi.org/10.1007/JHEP02(2016)052}
  {\path{doi:10.1007/JHEP02(2016)052}}.

\bibitem{deLeeuw:2016umh}
M.~de~Leeuw, C.~Kristjansen, S.~Mori, {AdS/dCFT one-point functions of the
  SU(3) sector}, Phys.Lett. B763 (2016) 197.
\newblock \href {http://arxiv.org/abs/1607.03123} {\path{arXiv:1607.03123}},
  \href {http://dx.doi.org/10.1016/j.physletb.2016.10.044}
  {\path{doi:10.1016/j.physletb.2016.10.044}}.

\bibitem{deLeeuw:2016ofj}
M.~de~Leeuw, C.~Kristjansen, G.~Linardopoulos, {One-point functions of
  non-protected operators in the SO(5) symmetric D3-D7 dCFT}, J.Phys. A50~(25)
  (2017) 254001.
\newblock \href {http://arxiv.org/abs/1612.06236} {\path{arXiv:1612.06236}},
  \href {http://dx.doi.org/10.1088/1751-8121/aa714b}
  {\path{doi:10.1088/1751-8121/aa714b}}.

\bibitem{deLeeuw:2018mkd}
M.~De~Leeuw, C.~Kristjansen, G.~Linardopoulos, {Scalar one-point functions and
  matrix product states of AdS/dCFT}, Phys. Lett. B781 (2018) 238--243.
\newblock \href {http://arxiv.org/abs/1802.01598} {\path{arXiv:1802.01598}},
  \href {http://dx.doi.org/10.1016/j.physletb.2018.03.083}
  {\path{doi:10.1016/j.physletb.2018.03.083}}.

\bibitem{Mestyan:2017xyk}
M.~{Mesty\'{a}n}, B.~Bertini, L.~Piroli, P.~Calabrese, {Exact solution for the
  quench dynamics of a nested integrable system}, J.Stat.Mech. 1708~(8) (2017)
  083103.
\newblock \href {http://arxiv.org/abs/1705.00851} {\path{arXiv:1705.00851}},
  \href {http://dx.doi.org/10.1088/1742-5468/aa7df0}
  {\path{doi:10.1088/1742-5468/aa7df0}}.

\bibitem{Piroli:2018ksf}
L.~Piroli, E.~Vernier, P.~Calabrese, B.~Pozsgay, {Integrable quenches in nested
  spin chains I: the exact steady states}\href
  {http://arxiv.org/abs/1811.00432} {\path{arXiv:1811.00432}}.

\bibitem{Piroli:2018don}
L.~Piroli, E.~Vernier, P.~Calabrese, B.~Pozsgay, {Integrable quenches in nested
  spin chains II: fusion of boundary transfer matrices}\href
  {http://arxiv.org/abs/1812.05330} {\path{arXiv:1812.05330}}.

\bibitem{Gaudin}
M.~Gaudin, {Diagonalisation d'une Classe d'Hamiltoniens de Spin}, J. Phys.
  France 37 (1976) 1087.

\bibitem{Korepin:1982ej}
V.~E. Korepin, {CALCULATION OF NORMS OF BETHE WAVE FUNCTIONS}, Commun. Math.
  Phys. 86 (1982) 391.

\bibitem{Piroli:2017sei}
L.~Piroli, B.~Pozsgay, E.~Vernier, {What is an integrable quench?}, Nucl.Phys.
  B925 (2017) 362.
\newblock \href {http://arxiv.org/abs/1709.04796} {\path{arXiv:1709.04796}},
  \href {http://dx.doi.org/10.1016/j.nuclphysb.2017.10.012}
  {\path{doi:10.1016/j.nuclphysb.2017.10.012}}.

\bibitem{Ghoshal:1993tm}
S.~Ghoshal, A.~B. Zamolodchikov, {Boundary S-matrix and boundary state in
  two-dimensional integrable quantum field theory}, Int.J.Mod.Phys. A9 (1994)
  3841, [Erratum: Int.J.Mod.Phys. A9 (1994) 4353].
\newblock \href {http://arxiv.org/abs/hep-th/9306002}
  {\path{arXiv:hep-th/9306002}}, \href
  {http://dx.doi.org/10.1142/S0217751X94001552}
  {\path{doi:10.1142/S0217751X94001552}}.

\bibitem{Pozsgay:2018dzs}
B.~Pozsgay, L.~Piroli, E.~Vernier, {Integrable Matrix Product States from
  boundary integrability}\href {http://arxiv.org/abs/1812.11094}
  {\path{arXiv:1812.11094}}.

\bibitem{Karch:2000gx}
A.~Karch, L.~Randall, {Open and closed string interpretation of SUSY CFT's on
  branes with boundaries}, JHEP 06 (2001) 063.
\newblock \href {http://arxiv.org/abs/hep-th/0105132}
  {\path{arXiv:hep-th/0105132}}, \href
  {http://dx.doi.org/10.1088/1126-6708/2001/06/063}
  {\path{doi:10.1088/1126-6708/2001/06/063}}.

\bibitem{Gaiotto:2008sa}
D.~Gaiotto, E.~Witten, {Supersymmetric Boundary Conditions in N=4 Super
  Yang-Mills Theory}, J. Statist. Phys. 135 (2009) 789--855.
\newblock \href {http://arxiv.org/abs/0804.2902} {\path{arXiv:0804.2902}},
  \href {http://dx.doi.org/10.1007/s10955-009-9687-3}
  {\path{doi:10.1007/s10955-009-9687-3}}.

\bibitem{Constable:1999ac}
N.~R. Constable, R.~C. Myers, O.~Tafjord, {The noncommutative bion core}, Phys.
  Rev. D61 (2000) 106009.
\newblock \href {http://arxiv.org/abs/hep-th/9911136}
  {\path{arXiv:hep-th/9911136}}, \href
  {http://dx.doi.org/10.1103/PhysRevD.61.106009}
  {\path{doi:10.1103/PhysRevD.61.106009}}.

\bibitem{Constable:2001ag}
N.~R. Constable, R.~C. Myers, O.~Tafjord, {Non-abelian brane intersections},
  JHEP 06 (2001) 023.
\newblock \href {http://arxiv.org/abs/hep-th/0102080}
  {\path{arXiv:hep-th/0102080}}, \href
  {http://dx.doi.org/10.1088/1126-6708/2001/06/023}
  {\path{doi:10.1088/1126-6708/2001/06/023}}.

\bibitem{Widen:2018nnu}
E.~Wid\'{e}n, {One-point functions in $\beta$-deformed $ \mathcal{N}=4 $ SYM
  with defect}, JHEP 11 (2018) 114.
\newblock \href {http://arxiv.org/abs/1804.09514} {\path{arXiv:1804.09514}},
  \href {http://dx.doi.org/10.1007/JHEP11(2018)114}
  {\path{doi:10.1007/JHEP11(2018)114}}.

\bibitem{Myers:2008me}
R.~C. Myers, M.~C. Wapler, {Transport Properties of Holographic Defects}, JHEP
  12 (2008) 115.
\newblock \href {http://arxiv.org/abs/0811.0480} {\path{arXiv:0811.0480}},
  \href {http://dx.doi.org/10.1088/1126-6708/2008/12/115}
  {\path{doi:10.1088/1126-6708/2008/12/115}}.

\bibitem{Bergman:2010gm}
O.~Bergman, N.~Jokela, G.~Lifschytz, M.~Lippert, {Quantum Hall Effect in a
  Holographic Model}, JHEP 10 (2010) 063.
\newblock \href {http://arxiv.org/abs/1003.4965} {\path{arXiv:1003.4965}},
  \href {http://dx.doi.org/10.1007/JHEP10(2010)063}
  {\path{doi:10.1007/JHEP10(2010)063}}.

\bibitem{Kristjansen:2012tn}
C.~Kristjansen, G.~W. Semenoff, D.~Young, {Chiral primary one-point functions
  in the D3-D7 defect conformal field theory}, JHEP 01 (2013) 117.
\newblock \href {http://arxiv.org/abs/1210.7015} {\path{arXiv:1210.7015}},
  \href {http://dx.doi.org/10.1007/JHEP01(2013)117}
  {\path{doi:10.1007/JHEP01(2013)117}}.

\bibitem{Jiang:2019xdz}
Y.~Jiang, S.~Komatsu, E.~Vescovi, {Structure Constants in $\mathcal{N}=4$ SYM
  at Finite Coupling as Worldsheet $g$-Function}\href
  {http://arxiv.org/abs/1906.07733} {\path{arXiv:1906.07733}}.

\bibitem{Bissi:2011dc}
A.~Bissi, C.~Kristjansen, D.~Young, K.~Zoubos, {Holographic three-point
  functions of giant gravitons}, JHEP 06 (2011) 085.
\newblock \href {http://arxiv.org/abs/1103.4079} {\path{arXiv:1103.4079}},
  \href {http://dx.doi.org/10.1007/JHEP06(2011)085}
  {\path{doi:10.1007/JHEP06(2011)085}}.

\bibitem{deLeeuw:2017cop}
M.~de~Leeuw, A.~C. Ipsen, C.~Kristjansen, M.~Wilhelm, {Introduction to
  integrability and one-point functions in $\mathcal{N}=4$ SYM and its defect
  cousin}, in: {Les Houches Summer School: Integrability: From Statistical
  Systems to Gauge Theory Les Houches, France, June 6-July 1, 2016}, 2017.
\newblock \href {http://arxiv.org/abs/1708.02525} {\path{arXiv:1708.02525}}.

\bibitem{Marius_review}
M.~de~Leeuw, {One-point functions in AdS/dCFT}, J. Phys. A: Math. Theor. in
  press.

\bibitem{Minahan:2002ve}
J.~A. Minahan, K.~Zarembo, {The Bethe ansatz for $\mathcal{N}=4$ super
  Yang-Mills}, JHEP 03 (2003) 013.
\newblock \href {http://arxiv.org/abs/hep-th/0212208}
  {\path{arXiv:hep-th/0212208}}, \href
  {http://dx.doi.org/10.1088/1126-6708/2003/03/013}
  {\path{doi:10.1088/1126-6708/2003/03/013}}.

\bibitem{Grau:2018keb}
A.~G. Grau, C.~Kristjansen, M.~Volk, M.~Wilhelm, {A Quantum Check of
  Non-Supersymmetric AdS/dCFT}, JHEP 01 (2019) 007.
\newblock \href {http://arxiv.org/abs/1810.11463} {\path{arXiv:1810.11463}},
  \href {http://dx.doi.org/10.1007/JHEP01(2019)007}
  {\path{doi:10.1007/JHEP01(2019)007}}.

\bibitem{Nagasaki:2012re}
K.~Nagasaki, S.~Yamaguchi, {Expectation values of chiral primary operators in
  holographic interface CFT}, Phys.Rev. D86 (2012) 086004.
\newblock \href {http://arxiv.org/abs/1205.1674} {\path{arXiv:1205.1674}},
  \href {http://dx.doi.org/10.1103/PhysRevD.86.086004}
  {\path{doi:10.1103/PhysRevD.86.086004}}.

\bibitem{Buhl-Mortensen:2016pxs}
I.~Buhl-Mortensen, M.~de~Leeuw, A.~C. Ipsen, C.~Kristjansen, M.~Wilhelm,
  {One-loop one-point functions in gauge-gravity dualities with defects}, Phys.
  Rev. Lett. 117~(23) (2016) 231603.
\newblock \href {http://arxiv.org/abs/1606.01886} {\path{arXiv:1606.01886}},
  \href {http://dx.doi.org/10.1103/PhysRevLett.117.231603}
  {\path{doi:10.1103/PhysRevLett.117.231603}}.

\bibitem{Buhl-Mortensen:2016jqo}
I.~Buhl-Mortensen, M.~de~Leeuw, A.~C. Ipsen, C.~Kristjansen, M.~Wilhelm, {A
  quantum check of AdS/dCFT}, JHEP 01 (2017) 098.
\newblock \href {http://arxiv.org/abs/1611.04603} {\path{arXiv:1611.04603}},
  \href {http://dx.doi.org/10.1007/JHEP01(2017)098}
  {\path{doi:10.1007/JHEP01(2017)098}}.

\bibitem{Buhl-Mortensen:2017ind}
I.~Buhl-Mortensen, M.~de~Leeuw, A.~C. Ipsen, C.~Kristjansen, M.~Wilhelm,
  {Asymptotic one-point functions in gauge-string duality with defects},
  Phys.Rev.Lett. 119~(26) (2017) 261604.
\newblock \href {http://arxiv.org/abs/1704.07386} {\path{arXiv:1704.07386}},
  \href {http://dx.doi.org/10.1103/PhysRevLett.119.261604}
  {\path{doi:10.1103/PhysRevLett.119.261604}}.

\end{thebibliography}




\end{document}